\documentstyle[12pt]{article}
\begin{document}
\begin{center}{\Large{\bf Monte Carlo study of dynamic phase transition 
in Ising metamagnet driven by oscillating magnetic field
}}\end{center}

\vskip 1cm

\begin{center}{\it Muktish Acharyya}\\
{\it Department of Physics, Presidency University}\\
{\it 86/1 College Street, Calcutta-700073, India}\\
{\it muktish.acharyya@gmail.com}\end{center}

\vskip 2cm

\noindent The dynamical responses of Ising metamagnet (layered
antiferromagnet) in the presence of a sinusoidally oscillating
magnetic field are studied by Monte Carlo simulation. 
The time
average staggered magnetization plays the role of dynamic order
parameter. A dynamical phase transition was observed and a phase
diagram was plotted in the plane formed by field amplitude and
temperature. The dynamical phase boundary is observed to shrink 
inward as the relative antiferromagnetic strength decreases. 
The results are compared with that obtained 
from pure ferromagnetic system. 
The shape of dynamic phase boundary observed to be 
qualitatively similar to that
obtained from previous meanfield calculations.

\vskip 2cm



\newpage

\noindent {\bf I. Introduction:}

The dynamic phase transition\cite{rmp},
in pure Ising ferromagnet driven by oscillating magnetic field,
became an interesting
field of modern research in nonequilibrium statistical physics. 
The researchers paid much attention in last few years 
to study the dynamical phase
transition in other magnetic models also. 
The dynamic transition is studied\cite{keskin}
in kinetic spin-3/2 Blume-Capel model.
The multicritical behaviour
was observed\cite{temizer} in kinetic Blume-Emery-Griffith model. 
The dynamic transition was investigated
in the classical Heisenberg model\cite{jung} with 
bilinear exchange anisotropy and in XY model\cite{xy}.
The multiple dynamic phase transition was also observed\cite{ma1}
in classical anisotropic Heisenberg ferromagnet
driven by polarised magnetic field. The existence of dynamic 
phase transition was found by few 
experimental studies in systems like ultrathin Co film
on Cu(100)\cite{ex1,ex2} by surface magneto-optic Kerr effect.

However all these studies mentioned above are done in simple ferromagnetic
systems. Due to the presence of complicated exchange interactions, 
the dynamical responses of metamagnets driven by oscillating 
magnetic field may give rise to some interesting effects. Keeping this
in mind, recently few researchers have taken interest to study the 
dynamical behaviours of metamagnets driven by oscillating magnetic field.
Few investigations are made in this front. Dynamic phase transition is 
studied in kinetic metamagnetic\cite{meta1} 
spin-3/2 Blume-Capel model, in Ising metamagnets\cite{meta2} etc. 
However, all these studies are mainly based on meanfield theory. 

In the meanfield calculations the spin fluctuations are ignored and the
results do not have any informations of microscopic details. 
Moreover, the transition is studied only from the temperature variations
of the order parameters. The temperature variations of quantity like
specific heat cannot be studied.
One possible
way to incorporate, the fluctuations as well as the temperature variations
of specific heat, is to study this by Monte Carlo
simulations. As far as the auhor's knowledge is concerned,
no such attempt has been made so far to study the dynamic phase transition,
even in simple Ising metamagnet driven by oscillating magnetic field,
by Monte Carlo simulation. Being motivated by these facts,
the dynamic response of Ising metamagnet 
driven by oscillating magnetic field, is studied
by Monte Carlo simulation, in this article.

The paper is organised in the following manner. The Ising metamagnetic model
and the Monte Carlo simulation scheme are discussed in the next section
(section-II). The numerical results are given in section-III. The paper ends
with concluding remarks and summary in section-IV.

\vskip 1cm

\noindent {\bf II. Model and Simulation:}

The time dependent Hamiltonian (or energy) of a three dimensional
Ising metamagnet (layered antiferromagnet) is
represented as:
\begin{equation}
H(t) = -{J_F}\Sigma_{F} s_i s_j -{J_A}\Sigma_{A} s_i s_j -h(t)\Sigma s_i.
\end{equation}
\noindent First term represents the in-plane ferromagnetic ($J_F > 0$)
nearest neighbour spin-spin 
interaction energy. Second term provides the interaction
energy coming from the antiferromagnetic ($J_A < 0$) interaction between 
two adjuscent layers
(since the nearest neighbour interactions are considered only). 
Third term gives the spin ($s_i = \pm 1$)-field
($h(t)$) interaction energy. Here the time dependent field is taken in the
sinusoidal form, i.e., $h(t) = h_0 {\rm cos}(\omega t)$. The boundary 
condition is taken periodic in all directions.

A cubic lattice of linear size $L=20$ is considered.
The dynamical evolution of the Ising spins are studied by Monte Carlo
simulation using Metropolis\cite{book}
single spin-flip scheme. The initial state of spin configuration
is prepared by taking 
50 percent of total number of spins 
(randomly selected)
up ($s_i=+1$). This corresponds to a high temperature paramagnetic phase.
According to Metropolis\cite{book} 
single spin flip dynamics, a spin, selected randomly, will flip ($s_i \to 
-s_i$) with probability,
\begin{equation}
W(s_i \to -s_i) = {\rm Min}[1,{\rm exp}(-\Delta H/{K_BT})]
\end{equation}
\noindent where, $\Delta H$ is the change in energy due to spin flip,
$K_B$ is Boltzmann constant and $T$ is the temperature. $L^3$ number 
of such random updates of spins is one Monte Carlo Step (MCS) and defines
the unit of time in the present study. After a long time a dynamical
steady state is achieved. The satisfactory steady values of the
time averages of the dynamical 
physical quantities (defined in next section)
ensures the achivement of steady state. 
All dynamical quantities  
are calculated in the steady state for a fixed
set of values of the temperature ($T$) and amplitude ($h_0$)
of the oscillating magnetic field.
Then the temperature is reduced (by a small step) keeping the values 
of other parameters unchanged and a similar process
(described above) is repeated. Here, the last spin configuration is
used as the initial state for the present value of temperature. In this
way the temperature variations of all dynamical quantities are studied.
Here, the temperature and field amplitudes are measured in the unit
where, $K_B=1$ and $J_F=1$.

\vskip 1cm
 
\noindent {\bf III. Numerical results:}

The zero temperature configuration of this system is all spins are
parallel in all layeres and the adjuscent layers contains antiparallel
spins. The system can be decomposed in two different sublattices
(say A and B). So,
alternate layeres form a sublattice. In a particular sublattice (A), the
instantaneous magnetization is $M_A(t) = 2(\Sigma_i s_i)/L^3$. The 
time average (over a full cycle of the oscillating magnetic field)
magnetization (for sublattice A) is 
$Q_A = {{\omega} \over {2\pi}}
\oint M_A(t) dt$. 
In the present study the frequency of the oscillating
magnetic field is taken $\omega = 2\pi f = 2\pi\times0.01$. 
The frequency ($f$) of the oscillating magnetic field is kept
constant ($f=0.01$) throughout the study.
So, one complete
oscillation would require 100 MCS. Initially, data for 1200 such cycles 
are discarded and average value of $Q_A$ is calculated from next 300
cycles. It is checked that 300 number of cycles are sufficient
to achieve the dynamically stable values of quantities. The dynamic order
parameter $Q_B$ for other sublattice (B) is calculated in the same way.
The staggered dynamic order parameter $Q_S$ is calculated as the
time average (over a full cycle of the oscillating magnetic field)
of instantaneous staggered magnetization (${{(M_A-M_B)} \over {2}}$). Here,  
$Q_S = {{\omega} \over {4\pi}}
\oint (M_A(t)-M_B(t)) dt$. The dynamical average energy is also
calculated as 
$E = {{\omega} \over {2\pi}} \oint H(t) dt$. 
The dynamical specific heat ($C$) is defined as $C={{dE} \over
{dT}}$ \cite{spht}.
The temperature variations (with step $\Delta T =0.05$)
of $Q_A$, $Q_B$, $Q_S$, $E$ and $C$ are studied here, 
considering the amplitude ($h_0$) and frequency ($f$)
of the time dependent magnetic field as parameters in each case.

Figure-1 shows the temperature variations of dynamic quantities for
fixed values of $J_F=1.0$, $J_A=-1.0$, $f=0.01$ and for two different
values ($h_0=2.5$ and $h_0=3.0$). Figure-1(a) shows the temperature
variations of the sublattice dynamic order parameters ($Q_A$ and $Q_B$).
This variations clearly shows the two dynamic transitions near 
$T_d = 2.2$ for $h_0=2.5$ and $T_d = 1.35$ for $h_0=3.0$.
One can easily visualize the temperature variations (not shown) of
dynamic staggered order parameters ($Q_S$).
These transitions becomes more pronounced from the plot of temperature
variations of the derivatives (${{dQ_S} \over {dT}}$) (in Figure-1(b)).
The sharp minima (eventually divergences in $L \to \infty$ limit) indicate
the dynamic transition temperatures $T_d=2.20$ and $T_d=1.35$ for 
$h_0=2.5$ and $h_0=3.0$ respectively. These two dynamic transition 
were reconfirmed and re-estimated {\it independently} 
from the studies of the 
temperature variations of dynamic specific
heat ($C$)\cite{spht}. It may be mentioned here that the temperature 
variation of the dynamic specific heat is not studied by 
previous meanfield
calculations\cite{meta2}
These variations are shown in Figure-1(c). Here also, the
dynamic specific heat ($C$) shows peaks (eventually divergences in
the limit $L\to \infty$) near $T_d=2.20$ and $T_d=1.35$ for $h_0=2.5$
and $h_0=3.0$ respectively. In this way, the transition temperatures
for the dynamic phase transitions are estimated over the range of values
of the amplitudes of the oscillating magnetic field. Thus for fixed values
of $J_A=-1.0$, $J_F=1.0$ and $f=0.01$, the dynamic phase boundary is
obtained.

To study the variations of dynamic phase boundary with the relative
strength of the antiferromagnetic interactions ($J_A$), the similar
investigations are made for $J_A=-0.5$ and $J_F=1.0$. 
In this case, the relative antiferromagnetic strength is reduced.
Figure-2(a)
shows the temperature variations of the derivative of staggered dynamic
order parameters, i.e., ${{dQ_S} \over {dT}}$. Like the previous cases,
the ${{dQ_S} \over {dT}}$, shows sharp minima indicating the dynamic
phase transitions near $T_d=1.20$ for $h_0=2.0$ and near $T_d=0.45$ for
$h_0=3.0$. The same transition temperatures were recalculted from the
temperature variations of dynmaic specific heat ($C$). The dynamic
specific heat $C$ shows sharp maxima near the same transition points
indicating the dynamic transitions. These results are shown in Figure-2(b).
In this way, the entire dynamic phase boundary is obtained.

For a comparison, the dynamic transitions are studied for a pure
ferromagnetic system. The transition temperatures $T_d=1.6$ 
and $T_d=0.8$
are obtained from the temperature variation of the derivative of
dynamic order parameter ${{dQ} \over {dT}}$, for $Q={\omega \over {2\pi}}
\oint M(t) dt$, where $M(t) = [M_A(t)+M_B(t)]$ 
is the total instantaneous magnetization obtained by putting
$J_F=1.0$ and $J_A=1.0$ in equation (1). The dynamic transition temperatures
are estimated from the sharp minima (shown in Figure-3) of 
${{dQ} \over {dT}}$. The results obtained here, are $T_d=1.6$ for $h_0
=2.0$ and $T_d=0.8$ for $h_0=3.0$. Hence the data for
entire dynamic phase
bounadry are obtained. 

The comprehensive results of the dynamic phase transitions are shown
as the dynamic phase boundary in Figure-4. From the phase diagrams
it is clear that the shapes of dynamic phase boundaries 
for metamagnetic dynamic
phase transitions are distinctly different from that for 
pure ferromagnets. Additionally, the dynamic phase boundary shrinks
inward as the relative strength of antiferromagnetic interaction decreases.
These phase diagrams for metamagnetic dynamic phase transitions agrees
qualitatively well with those obtained from meanfield 
calculations\cite{meta2}. 

\vskip 1cm

\noindent {\bf IV. Concluding remarks:}

The dynamic phase transition in Ising metamagnet driven by 
oscillating magnetic field is studied by Monte Carlo simulation. As far
as the author's knowledge is concerned, this is the first Monte Carlo
study of dynamic phase transition for this type of metamagnetic model.
This study differes from the previous meanfield study\cite{meta2} in the
following aspects: {\it Firstly,} 
the dynamic phase transition was reconfirmed
here from the temperature variation of dynamic specific heat\cite{spht}
, which was not done in meanfield study\cite{meta2}. 
{\it Secondly,} in meanfield
study, the multicritical behaviours are observed and the tricritical point
was located on the phase boundary at lower temperatures. Here, in the 
Monte Carlo study, the probability of spin flip was calculated from 
the Gibb's
distribution which provides good results in high temperature region. In
the low temperature, region the quantum fluctuation will be excited and
Quantum Monte Carlo will be the better method of study. For this reason,
the phase boundary for very low temperature is not drawn. 

In the present
study, the order of transition is not mentioned. As a result one cannot
be able to guess about the multicritical behaviour. Here, only the dynamic
{\it transition} is observed. It may be mentioned here, that the 
existence of tricritical
point was found\cite{ma2} 
on the dynamic phase boundary by Monte Carlo simulation
in pure Ising ferromagnet. However, this was disproved\cite{korniss}
later from the
Monte Carlo study with larger lattice size and improved statistics.
So, it is better not to mention about multicritical behaviour from the
present Monte Carlo study of such small system size ($L=20$). 

The shapes of the dynamic phase boundary depend on the relative 
antiferromagnetic strength as well the the frequencies of the oscillating
magnetic field. In the present case, this variation with frequency
is not studied. It may be an interesting study.

Above all, this dynamic phase transition in Ising metamagnets 
gives interesting result, theoretically. The physical reason
of the dynamic phase transition in metamagnet may be stated 
qualitatively as
follows: for small values of the amplitude of the field and at low 
temperature, the instantaneous staggered magnetization oscillates
about a nonzero value, giving rise to ordered phase. However, for
large value of field amplitude and at high temperature, the instantaneous
staggered magnetization oscillates symmetrically about a value very
close to zero, leading to disordered phase.
One has to see the hysteretic loop ($m-h$) \cite{rmp}
to get the clear idea about the physical 
interpretation of the dynamic phase transition. The asymmetric
$m-h$ loop gives the dynamically ordered phase and the symmetric
$m-h$ loop gives disordered phase. In the case of metamagnet, this
$m$ should be the staggered magnetization. 
This theoretical investigation
is an apeal to the experimentalists to study the dynamic phase transition
in anisotropic metamagnet, like ${\rm FeBr_2}$. 
Experimentally the existence of dynamic transition was found
\cite{ex1,ex2} in Co film (ferromagnetic)
on Cu surface (at room temperature)
 by surface magneto optic Kerr effect. 
Let us hope that the experimental study
of dynamic phase transition in metamagnetic systems like ${\rm FeBr_2}$
driven by oscillating magnetic field, 
will bring some new physics in future. 

\vskip 1cm

\noindent {\bf Acknowledgements:} The library facilities provided by 
Calcutta University is gratefully acknowledged. Author would also like
to thank Prabir Banerjee, for helping him to prepare the manuscript.

\newpage
\setlength{\unitlength}{0.240900pt}
\ifx\plotpoint\undefined\newsavebox{\plotpoint}\fi
\sbox{\plotpoint}{\rule[-0.200pt]{0.400pt}{0.400pt}}%


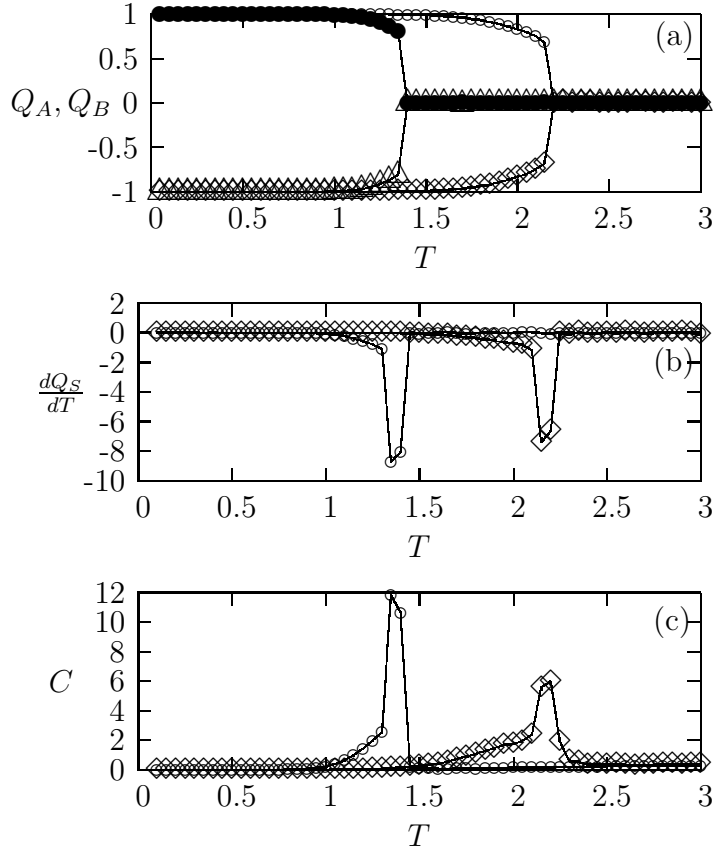
\noindent {\bf Fig-1.} The temperature variations of (a) sublattice
dynamic order parameters $Q_A$ (($\bullet$) for
$h_0=3.0$ and (o) for $h_0=2.5$)and $Q_B$
(($\triangle$) for $h_0=3.0$ and ($\diamond$) for $h_0=2.5$), 
(b) the derivative (${{dQ_S} \over {dT}}$)
of staggered dynamic order parameter ($Q_S$)
((o) for $h_0=3.0$ and ($\diamond$) for $h_0=2.5$)
 and (c) the
dynamic specific heat ($C$)
((o) for $h_0=3.0$ and ($\diamond$) for $h_0=2.5$).
The continuous lines in all cases joining the data points act
as guides to the eye.
Here, $J_A=-1\times J_F$.
The $T_d(h_0=2.5)=2.2$ and
$T_d(h_0=3.0)=1.35$.
\newpage
\setlength{\unitlength}{0.240900pt}
\ifx\plotpoint\undefined\newsavebox{\plotpoint}\fi
\sbox{\plotpoint}{\rule[-0.200pt]{0.400pt}{0.400pt}}%
\begin{picture}(1125,450)(0,0)
\sbox{\plotpoint}{\rule[-0.200pt]{0.400pt}{0.400pt}}%
\put(201.0,131.0){\rule[-0.200pt]{4.818pt}{0.400pt}}
\put(181,131){\makebox(0,0)[r]{-10}}
\put(1064.0,131.0){\rule[-0.200pt]{4.818pt}{0.400pt}}
\put(201.0,178.0){\rule[-0.200pt]{4.818pt}{0.400pt}}
\put(181,178){\makebox(0,0)[r]{-8}}
\put(1064.0,178.0){\rule[-0.200pt]{4.818pt}{0.400pt}}
\put(201.0,224.0){\rule[-0.200pt]{4.818pt}{0.400pt}}
\put(181,224){\makebox(0,0)[r]{-6}}
\put(1064.0,224.0){\rule[-0.200pt]{4.818pt}{0.400pt}}
\put(201.0,271.0){\rule[-0.200pt]{4.818pt}{0.400pt}}
\put(181,271){\makebox(0,0)[r]{-4}}
\put(1064.0,271.0){\rule[-0.200pt]{4.818pt}{0.400pt}}
\put(201.0,317.0){\rule[-0.200pt]{4.818pt}{0.400pt}}
\put(181,317){\makebox(0,0)[r]{-2}}
\put(1064.0,317.0){\rule[-0.200pt]{4.818pt}{0.400pt}}
\put(201.0,364.0){\rule[-0.200pt]{4.818pt}{0.400pt}}
\put(181,364){\makebox(0,0)[r]{ 0}}
\put(1064.0,364.0){\rule[-0.200pt]{4.818pt}{0.400pt}}
\put(201.0,410.0){\rule[-0.200pt]{4.818pt}{0.400pt}}
\put(181,410){\makebox(0,0)[r]{ 2}}
\put(1064.0,410.0){\rule[-0.200pt]{4.818pt}{0.400pt}}
\put(201.0,131.0){\rule[-0.200pt]{0.400pt}{4.818pt}}
\put(201,90){\makebox(0,0){ 0}}
\put(201.0,390.0){\rule[-0.200pt]{0.400pt}{4.818pt}}
\put(422.0,131.0){\rule[-0.200pt]{0.400pt}{4.818pt}}
\put(422,90){\makebox(0,0){ 0.5}}
\put(422.0,390.0){\rule[-0.200pt]{0.400pt}{4.818pt}}
\put(643.0,131.0){\rule[-0.200pt]{0.400pt}{4.818pt}}
\put(643,90){\makebox(0,0){ 1}}
\put(643.0,390.0){\rule[-0.200pt]{0.400pt}{4.818pt}}
\put(863.0,131.0){\rule[-0.200pt]{0.400pt}{4.818pt}}
\put(863,90){\makebox(0,0){ 1.5}}
\put(863.0,390.0){\rule[-0.200pt]{0.400pt}{4.818pt}}
\put(1084.0,131.0){\rule[-0.200pt]{0.400pt}{4.818pt}}
\put(1084,90){\makebox(0,0){ 2}}
\put(1084.0,390.0){\rule[-0.200pt]{0.400pt}{4.818pt}}
\put(201.0,131.0){\rule[-0.200pt]{0.400pt}{67.211pt}}
\put(201.0,131.0){\rule[-0.200pt]{212.715pt}{0.400pt}}
\put(1084.0,131.0){\rule[-0.200pt]{0.400pt}{67.211pt}}
\put(201.0,410.0){\rule[-0.200pt]{212.715pt}{0.400pt}}
\put(80,270){\makebox(0,0){${{dQ_S} \over {dT}}$}}
\put(642,29){\makebox(0,0){$T$}}
\put(974,317){\makebox(0,0)[l]{(a)}}
\put(1084,363){\raisebox{-.8pt}{\makebox(0,0){$\Diamond$}}}
\put(1062,363){\raisebox{-.8pt}{\makebox(0,0){$\Diamond$}}}
\put(1040,363){\raisebox{-.8pt}{\makebox(0,0){$\Diamond$}}}
\put(1018,364){\raisebox{-.8pt}{\makebox(0,0){$\Diamond$}}}
\put(996,364){\raisebox{-.8pt}{\makebox(0,0){$\Diamond$}}}
\put(974,363){\raisebox{-.8pt}{\makebox(0,0){$\Diamond$}}}
\put(952,363){\raisebox{-.8pt}{\makebox(0,0){$\Diamond$}}}
\put(929,364){\raisebox{-.8pt}{\makebox(0,0){$\Diamond$}}}
\put(907,364){\raisebox{-.8pt}{\makebox(0,0){$\Diamond$}}}
\put(885,363){\raisebox{-.8pt}{\makebox(0,0){$\Diamond$}}}
\put(863,363){\raisebox{-.8pt}{\makebox(0,0){$\Diamond$}}}
\put(841,363){\raisebox{-.8pt}{\makebox(0,0){$\Diamond$}}}
\put(819,364){\raisebox{-.8pt}{\makebox(0,0){$\Diamond$}}}
\put(797,364){\raisebox{-.8pt}{\makebox(0,0){$\Diamond$}}}
\put(775,355){\raisebox{-.8pt}{\makebox(0,0){$\Diamond$}}}
\put(753,159){\raisebox{-.8pt}{\makebox(0,0){$\Diamond$}}}
\put(731,155){\raisebox{-.8pt}{\makebox(0,0){$\Diamond$}}}
\put(709,344){\raisebox{-.8pt}{\makebox(0,0){$\Diamond$}}}
\put(687,352){\raisebox{-.8pt}{\makebox(0,0){$\Diamond$}}}
\put(665,357){\raisebox{-.8pt}{\makebox(0,0){$\Diamond$}}}
\put(643,360){\raisebox{-.8pt}{\makebox(0,0){$\Diamond$}}}
\put(620,362){\raisebox{-.8pt}{\makebox(0,0){$\Diamond$}}}
\put(598,363){\raisebox{-.8pt}{\makebox(0,0){$\Diamond$}}}
\put(576,363){\raisebox{-.8pt}{\makebox(0,0){$\Diamond$}}}
\put(554,363){\raisebox{-.8pt}{\makebox(0,0){$\Diamond$}}}
\put(532,363){\raisebox{-.8pt}{\makebox(0,0){$\Diamond$}}}
\put(510,363){\raisebox{-.8pt}{\makebox(0,0){$\Diamond$}}}
\put(488,363){\raisebox{-.8pt}{\makebox(0,0){$\Diamond$}}}
\put(466,363){\raisebox{-.8pt}{\makebox(0,0){$\Diamond$}}}
\put(444,363){\raisebox{-.8pt}{\makebox(0,0){$\Diamond$}}}
\put(422,363){\raisebox{-.8pt}{\makebox(0,0){$\Diamond$}}}
\put(400,363){\raisebox{-.8pt}{\makebox(0,0){$\Diamond$}}}
\put(378,363){\raisebox{-.8pt}{\makebox(0,0){$\Diamond$}}}
\put(356,363){\raisebox{-.8pt}{\makebox(0,0){$\Diamond$}}}
\put(333,364){\raisebox{-.8pt}{\makebox(0,0){$\Diamond$}}}
\put(311,364){\raisebox{-.8pt}{\makebox(0,0){$\Diamond$}}}
\put(289,364){\raisebox{-.8pt}{\makebox(0,0){$\Diamond$}}}
\put(267,364){\raisebox{-.8pt}{\makebox(0,0){$\Diamond$}}}
\put(245,364){\raisebox{-.8pt}{\makebox(0,0){$\Diamond$}}}
\put(1084,363){\circle{18}}
\put(1062,363){\circle{18}}
\put(1040,363){\circle{18}}
\put(1018,364){\circle{18}}
\put(996,364){\circle{18}}
\put(974,364){\circle{18}}
\put(952,363){\circle{18}}
\put(929,363){\circle{18}}
\put(907,364){\circle{18}}
\put(885,364){\circle{18}}
\put(863,364){\circle{18}}
\put(841,363){\circle{18}}
\put(819,363){\circle{18}}
\put(797,363){\circle{18}}
\put(775,363){\circle{18}}
\put(753,363){\circle{18}}
\put(731,364){\circle{18}}
\put(709,364){\circle{18}}
\put(687,363){\circle{18}}
\put(665,364){\circle{18}}
\put(643,364){\circle{18}}
\put(620,363){\circle{18}}
\put(598,363){\circle{18}}
\put(576,363){\circle{18}}
\put(554,364){\circle{18}}
\put(532,364){\circle{18}}
\put(510,363){\circle{18}}
\put(488,364){\circle{18}}
\put(466,362){\circle{18}}
\put(444,364){\circle{18}}
\put(422,212){\circle{18}}
\put(400,143){\circle{18}}
\put(378,287){\circle{18}}
\put(356,352){\circle{18}}
\put(333,361){\circle{18}}
\put(311,363){\circle{18}}
\put(289,363){\circle{18}}
\put(267,364){\circle{18}}
\put(245,364){\circle{18}}
\put(1084,363){\usebox{\plotpoint}}
\put(1018,362.67){\rule{5.300pt}{0.400pt}}
\multiput(1029.00,362.17)(-11.000,1.000){2}{\rule{2.650pt}{0.400pt}}
\put(1040.0,363.0){\rule[-0.200pt]{10.600pt}{0.400pt}}
\put(974,362.67){\rule{5.300pt}{0.400pt}}
\multiput(985.00,363.17)(-11.000,-1.000){2}{\rule{2.650pt}{0.400pt}}
\put(996.0,364.0){\rule[-0.200pt]{5.300pt}{0.400pt}}
\put(929,362.67){\rule{5.541pt}{0.400pt}}
\multiput(940.50,362.17)(-11.500,1.000){2}{\rule{2.770pt}{0.400pt}}
\put(952.0,363.0){\rule[-0.200pt]{5.300pt}{0.400pt}}
\put(885,362.67){\rule{5.300pt}{0.400pt}}
\multiput(896.00,363.17)(-11.000,-1.000){2}{\rule{2.650pt}{0.400pt}}
\put(907.0,364.0){\rule[-0.200pt]{5.300pt}{0.400pt}}
\put(819,362.67){\rule{5.300pt}{0.400pt}}
\multiput(830.00,362.17)(-11.000,1.000){2}{\rule{2.650pt}{0.400pt}}
\put(841.0,363.0){\rule[-0.200pt]{10.600pt}{0.400pt}}
\multiput(792.53,362.93)(-1.252,-0.489){15}{\rule{1.078pt}{0.118pt}}
\multiput(794.76,363.17)(-19.763,-9.000){2}{\rule{0.539pt}{0.400pt}}
\multiput(773.92,339.79)(-0.496,-4.520){41}{\rule{0.120pt}{3.664pt}}
\multiput(774.17,347.40)(-22.000,-188.396){2}{\rule{0.400pt}{1.832pt}}
\multiput(743.45,157.94)(-3.113,-0.468){5}{\rule{2.300pt}{0.113pt}}
\multiput(748.23,158.17)(-17.226,-4.000){2}{\rule{1.150pt}{0.400pt}}
\multiput(729.92,155.00)(-0.496,4.358){41}{\rule{0.120pt}{3.536pt}}
\multiput(730.17,155.00)(-22.000,181.660){2}{\rule{0.400pt}{1.768pt}}
\multiput(704.02,344.59)(-1.418,0.488){13}{\rule{1.200pt}{0.117pt}}
\multiput(706.51,343.17)(-19.509,8.000){2}{\rule{0.600pt}{0.400pt}}
\multiput(679.28,352.59)(-2.380,0.477){7}{\rule{1.860pt}{0.115pt}}
\multiput(683.14,351.17)(-18.139,5.000){2}{\rule{0.930pt}{0.400pt}}
\multiput(652.41,357.61)(-4.704,0.447){3}{\rule{3.033pt}{0.108pt}}
\multiput(658.70,356.17)(-15.704,3.000){2}{\rule{1.517pt}{0.400pt}}
\put(620,360.17){\rule{4.700pt}{0.400pt}}
\multiput(633.24,359.17)(-13.245,2.000){2}{\rule{2.350pt}{0.400pt}}
\put(598,361.67){\rule{5.300pt}{0.400pt}}
\multiput(609.00,361.17)(-11.000,1.000){2}{\rule{2.650pt}{0.400pt}}
\put(797.0,364.0){\rule[-0.200pt]{5.300pt}{0.400pt}}
\put(333,362.67){\rule{5.541pt}{0.400pt}}
\multiput(344.50,362.17)(-11.500,1.000){2}{\rule{2.770pt}{0.400pt}}
\put(356.0,363.0){\rule[-0.200pt]{58.298pt}{0.400pt}}
\put(245.0,364.0){\rule[-0.200pt]{21.199pt}{0.400pt}}
\put(1084,363){\usebox{\plotpoint}}
\put(1018,362.67){\rule{5.300pt}{0.400pt}}
\multiput(1029.00,362.17)(-11.000,1.000){2}{\rule{2.650pt}{0.400pt}}
\put(1040.0,363.0){\rule[-0.200pt]{10.600pt}{0.400pt}}
\put(952,362.67){\rule{5.300pt}{0.400pt}}
\multiput(963.00,363.17)(-11.000,-1.000){2}{\rule{2.650pt}{0.400pt}}
\put(974.0,364.0){\rule[-0.200pt]{10.600pt}{0.400pt}}
\put(907,362.67){\rule{5.300pt}{0.400pt}}
\multiput(918.00,362.17)(-11.000,1.000){2}{\rule{2.650pt}{0.400pt}}
\put(929.0,363.0){\rule[-0.200pt]{5.541pt}{0.400pt}}
\put(841,362.67){\rule{5.300pt}{0.400pt}}
\multiput(852.00,363.17)(-11.000,-1.000){2}{\rule{2.650pt}{0.400pt}}
\put(863.0,364.0){\rule[-0.200pt]{10.600pt}{0.400pt}}
\put(731,362.67){\rule{5.300pt}{0.400pt}}
\multiput(742.00,362.17)(-11.000,1.000){2}{\rule{2.650pt}{0.400pt}}
\put(753.0,363.0){\rule[-0.200pt]{21.199pt}{0.400pt}}
\put(687,362.67){\rule{5.300pt}{0.400pt}}
\multiput(698.00,363.17)(-11.000,-1.000){2}{\rule{2.650pt}{0.400pt}}
\put(665,362.67){\rule{5.300pt}{0.400pt}}
\multiput(676.00,362.17)(-11.000,1.000){2}{\rule{2.650pt}{0.400pt}}
\put(709.0,364.0){\rule[-0.200pt]{5.300pt}{0.400pt}}
\put(620,362.67){\rule{5.541pt}{0.400pt}}
\multiput(631.50,363.17)(-11.500,-1.000){2}{\rule{2.770pt}{0.400pt}}
\put(643.0,364.0){\rule[-0.200pt]{5.300pt}{0.400pt}}
\put(554,362.67){\rule{5.300pt}{0.400pt}}
\multiput(565.00,362.17)(-11.000,1.000){2}{\rule{2.650pt}{0.400pt}}
\put(576.0,363.0){\rule[-0.200pt]{10.600pt}{0.400pt}}
\put(510,362.67){\rule{5.300pt}{0.400pt}}
\multiput(521.00,363.17)(-11.000,-1.000){2}{\rule{2.650pt}{0.400pt}}
\put(488,362.67){\rule{5.300pt}{0.400pt}}
\multiput(499.00,362.17)(-11.000,1.000){2}{\rule{2.650pt}{0.400pt}}
\put(466,362.17){\rule{4.500pt}{0.400pt}}
\multiput(478.66,363.17)(-12.660,-2.000){2}{\rule{2.250pt}{0.400pt}}
\put(444,362.17){\rule{4.500pt}{0.400pt}}
\multiput(456.66,361.17)(-12.660,2.000){2}{\rule{2.250pt}{0.400pt}}
\multiput(442.92,352.11)(-0.496,-3.503){41}{\rule{0.120pt}{2.864pt}}
\multiput(443.17,358.06)(-22.000,-146.056){2}{\rule{0.400pt}{1.432pt}}
\multiput(420.92,206.38)(-0.496,-1.584){41}{\rule{0.120pt}{1.355pt}}
\multiput(421.17,209.19)(-22.000,-66.189){2}{\rule{0.400pt}{0.677pt}}
\multiput(398.92,143.00)(-0.496,3.318){41}{\rule{0.120pt}{2.718pt}}
\multiput(399.17,143.00)(-22.000,138.358){2}{\rule{0.400pt}{1.359pt}}
\multiput(376.92,287.00)(-0.496,1.492){41}{\rule{0.120pt}{1.282pt}}
\multiput(377.17,287.00)(-22.000,62.340){2}{\rule{0.400pt}{0.641pt}}
\multiput(351.34,352.59)(-1.310,0.489){15}{\rule{1.122pt}{0.118pt}}
\multiput(353.67,351.17)(-20.671,9.000){2}{\rule{0.561pt}{0.400pt}}
\put(311,361.17){\rule{4.500pt}{0.400pt}}
\multiput(323.66,360.17)(-12.660,2.000){2}{\rule{2.250pt}{0.400pt}}
\put(532.0,364.0){\rule[-0.200pt]{5.300pt}{0.400pt}}
\put(267,362.67){\rule{5.300pt}{0.400pt}}
\multiput(278.00,362.17)(-11.000,1.000){2}{\rule{2.650pt}{0.400pt}}
\put(289.0,363.0){\rule[-0.200pt]{5.300pt}{0.400pt}}
\put(245.0,364.0){\rule[-0.200pt]{5.300pt}{0.400pt}}
\put(201.0,131.0){\rule[-0.200pt]{0.400pt}{67.211pt}}
\put(201.0,131.0){\rule[-0.200pt]{212.715pt}{0.400pt}}
\put(1084.0,131.0){\rule[-0.200pt]{0.400pt}{67.211pt}}
\put(201.0,410.0){\rule[-0.200pt]{212.715pt}{0.400pt}}
\end{picture}

\setlength{\unitlength}{0.240900pt}
\ifx\plotpoint\undefined\newsavebox{\plotpoint}\fi
\sbox{\plotpoint}{\rule[-0.200pt]{0.400pt}{0.400pt}}%
\begin{picture}(1125,450)(0,0)
\sbox{\plotpoint}{\rule[-0.200pt]{0.400pt}{0.400pt}}%
\put(201.0,131.0){\rule[-0.200pt]{4.818pt}{0.400pt}}
\put(181,131){\makebox(0,0)[r]{ 0}}
\put(1064.0,131.0){\rule[-0.200pt]{4.818pt}{0.400pt}}
\put(201.0,187.0){\rule[-0.200pt]{4.818pt}{0.400pt}}
\put(181,187){\makebox(0,0)[r]{ 2}}
\put(1064.0,187.0){\rule[-0.200pt]{4.818pt}{0.400pt}}
\put(201.0,243.0){\rule[-0.200pt]{4.818pt}{0.400pt}}
\put(181,243){\makebox(0,0)[r]{ 4}}
\put(1064.0,243.0){\rule[-0.200pt]{4.818pt}{0.400pt}}
\put(201.0,298.0){\rule[-0.200pt]{4.818pt}{0.400pt}}
\put(181,298){\makebox(0,0)[r]{ 6}}
\put(1064.0,298.0){\rule[-0.200pt]{4.818pt}{0.400pt}}
\put(201.0,354.0){\rule[-0.200pt]{4.818pt}{0.400pt}}
\put(181,354){\makebox(0,0)[r]{ 8}}
\put(1064.0,354.0){\rule[-0.200pt]{4.818pt}{0.400pt}}
\put(201.0,410.0){\rule[-0.200pt]{4.818pt}{0.400pt}}
\put(181,410){\makebox(0,0)[r]{ 10}}
\put(1064.0,410.0){\rule[-0.200pt]{4.818pt}{0.400pt}}
\put(201.0,131.0){\rule[-0.200pt]{0.400pt}{4.818pt}}
\put(201,90){\makebox(0,0){ 0}}
\put(201.0,390.0){\rule[-0.200pt]{0.400pt}{4.818pt}}
\put(422.0,131.0){\rule[-0.200pt]{0.400pt}{4.818pt}}
\put(422,90){\makebox(0,0){ 0.5}}
\put(422.0,390.0){\rule[-0.200pt]{0.400pt}{4.818pt}}
\put(643.0,131.0){\rule[-0.200pt]{0.400pt}{4.818pt}}
\put(643,90){\makebox(0,0){ 1}}
\put(643.0,390.0){\rule[-0.200pt]{0.400pt}{4.818pt}}
\put(863.0,131.0){\rule[-0.200pt]{0.400pt}{4.818pt}}
\put(863,90){\makebox(0,0){ 1.5}}
\put(863.0,390.0){\rule[-0.200pt]{0.400pt}{4.818pt}}
\put(1084.0,131.0){\rule[-0.200pt]{0.400pt}{4.818pt}}
\put(1084,90){\makebox(0,0){ 2}}
\put(1084.0,390.0){\rule[-0.200pt]{0.400pt}{4.818pt}}
\put(201.0,131.0){\rule[-0.200pt]{0.400pt}{67.211pt}}
\put(201.0,131.0){\rule[-0.200pt]{212.715pt}{0.400pt}}
\put(1084.0,131.0){\rule[-0.200pt]{0.400pt}{67.211pt}}
\put(201.0,410.0){\rule[-0.200pt]{212.715pt}{0.400pt}}
\put(80,270){\makebox(0,0){$C$}}
\put(642,29){\makebox(0,0){$T$}}
\put(974,382){\makebox(0,0)[l]{(b)}}
\put(1084,137){\raisebox{-.8pt}{\makebox(0,0){$\Diamond$}}}
\put(1062,136){\raisebox{-.8pt}{\makebox(0,0){$\Diamond$}}}
\put(1040,136){\raisebox{-.8pt}{\makebox(0,0){$\Diamond$}}}
\put(1018,136){\raisebox{-.8pt}{\makebox(0,0){$\Diamond$}}}
\put(996,135){\raisebox{-.8pt}{\makebox(0,0){$\Diamond$}}}
\put(974,135){\raisebox{-.8pt}{\makebox(0,0){$\Diamond$}}}
\put(952,135){\raisebox{-.8pt}{\makebox(0,0){$\Diamond$}}}
\put(929,134){\raisebox{-.8pt}{\makebox(0,0){$\Diamond$}}}
\put(907,134){\raisebox{-.8pt}{\makebox(0,0){$\Diamond$}}}
\put(885,134){\raisebox{-.8pt}{\makebox(0,0){$\Diamond$}}}
\put(863,133){\raisebox{-.8pt}{\makebox(0,0){$\Diamond$}}}
\put(841,133){\raisebox{-.8pt}{\makebox(0,0){$\Diamond$}}}
\put(819,134){\raisebox{-.8pt}{\makebox(0,0){$\Diamond$}}}
\put(797,133){\raisebox{-.8pt}{\makebox(0,0){$\Diamond$}}}
\put(775,141){\raisebox{-.8pt}{\makebox(0,0){$\Diamond$}}}
\put(753,352){\raisebox{-.8pt}{\makebox(0,0){$\Diamond$}}}
\put(731,364){\raisebox{-.8pt}{\makebox(0,0){$\Diamond$}}}
\put(709,167){\raisebox{-.8pt}{\makebox(0,0){$\Diamond$}}}
\put(687,154){\raisebox{-.8pt}{\makebox(0,0){$\Diamond$}}}
\put(665,145){\raisebox{-.8pt}{\makebox(0,0){$\Diamond$}}}
\put(643,139){\raisebox{-.8pt}{\makebox(0,0){$\Diamond$}}}
\put(620,136){\raisebox{-.8pt}{\makebox(0,0){$\Diamond$}}}
\put(598,134){\raisebox{-.8pt}{\makebox(0,0){$\Diamond$}}}
\put(576,132){\raisebox{-.8pt}{\makebox(0,0){$\Diamond$}}}
\put(554,132){\raisebox{-.8pt}{\makebox(0,0){$\Diamond$}}}
\put(532,131){\raisebox{-.8pt}{\makebox(0,0){$\Diamond$}}}
\put(510,131){\raisebox{-.8pt}{\makebox(0,0){$\Diamond$}}}
\put(488,131){\raisebox{-.8pt}{\makebox(0,0){$\Diamond$}}}
\put(466,131){\raisebox{-.8pt}{\makebox(0,0){$\Diamond$}}}
\put(444,131){\raisebox{-.8pt}{\makebox(0,0){$\Diamond$}}}
\put(422,131){\raisebox{-.8pt}{\makebox(0,0){$\Diamond$}}}
\put(400,131){\raisebox{-.8pt}{\makebox(0,0){$\Diamond$}}}
\put(378,131){\raisebox{-.8pt}{\makebox(0,0){$\Diamond$}}}
\put(356,131){\raisebox{-.8pt}{\makebox(0,0){$\Diamond$}}}
\put(333,131){\raisebox{-.8pt}{\makebox(0,0){$\Diamond$}}}
\put(311,131){\raisebox{-.8pt}{\makebox(0,0){$\Diamond$}}}
\put(289,131){\raisebox{-.8pt}{\makebox(0,0){$\Diamond$}}}
\put(267,131){\raisebox{-.8pt}{\makebox(0,0){$\Diamond$}}}
\put(245,131){\raisebox{-.8pt}{\makebox(0,0){$\Diamond$}}}
\put(1084,135){\circle{18}}
\put(1062,134){\circle{18}}
\put(1040,134){\circle{18}}
\put(1018,134){\circle{18}}
\put(996,134){\circle{18}}
\put(974,134){\circle{18}}
\put(952,134){\circle{18}}
\put(929,133){\circle{18}}
\put(907,133){\circle{18}}
\put(885,133){\circle{18}}
\put(863,133){\circle{18}}
\put(841,133){\circle{18}}
\put(819,133){\circle{18}}
\put(797,133){\circle{18}}
\put(775,133){\circle{18}}
\put(753,133){\circle{18}}
\put(731,133){\circle{18}}
\put(709,133){\circle{18}}
\put(687,133){\circle{18}}
\put(665,133){\circle{18}}
\put(643,133){\circle{18}}
\put(620,133){\circle{18}}
\put(598,133){\circle{18}}
\put(576,133){\circle{18}}
\put(554,133){\circle{18}}
\put(532,133){\circle{18}}
\put(510,133){\circle{18}}
\put(488,133){\circle{18}}
\put(466,135){\circle{18}}
\put(444,158){\circle{18}}
\put(422,310){\circle{18}}
\put(400,361){\circle{18}}
\put(378,220){\circle{18}}
\put(356,147){\circle{18}}
\put(333,134){\circle{18}}
\put(311,131){\circle{18}}
\put(289,131){\circle{18}}
\put(267,131){\circle{18}}
\put(245,131){\circle{18}}
\put(1084,137){\usebox{\plotpoint}}
\put(1062,135.67){\rule{5.300pt}{0.400pt}}
\multiput(1073.00,136.17)(-11.000,-1.000){2}{\rule{2.650pt}{0.400pt}}
\put(996,134.67){\rule{5.300pt}{0.400pt}}
\multiput(1007.00,135.17)(-11.000,-1.000){2}{\rule{2.650pt}{0.400pt}}
\put(1018.0,136.0){\rule[-0.200pt]{10.600pt}{0.400pt}}
\put(929,133.67){\rule{5.541pt}{0.400pt}}
\multiput(940.50,134.17)(-11.500,-1.000){2}{\rule{2.770pt}{0.400pt}}
\put(952.0,135.0){\rule[-0.200pt]{10.600pt}{0.400pt}}
\put(863,132.67){\rule{5.300pt}{0.400pt}}
\multiput(874.00,133.17)(-11.000,-1.000){2}{\rule{2.650pt}{0.400pt}}
\put(885.0,134.0){\rule[-0.200pt]{10.600pt}{0.400pt}}
\put(819,132.67){\rule{5.300pt}{0.400pt}}
\multiput(830.00,132.17)(-11.000,1.000){2}{\rule{2.650pt}{0.400pt}}
\put(797,132.67){\rule{5.300pt}{0.400pt}}
\multiput(808.00,133.17)(-11.000,-1.000){2}{\rule{2.650pt}{0.400pt}}
\multiput(792.02,133.59)(-1.418,0.488){13}{\rule{1.200pt}{0.117pt}}
\multiput(794.51,132.17)(-19.509,8.000){2}{\rule{0.600pt}{0.400pt}}
\multiput(773.92,141.00)(-0.496,4.866){41}{\rule{0.120pt}{3.936pt}}
\multiput(774.17,141.00)(-22.000,202.830){2}{\rule{0.400pt}{1.968pt}}
\multiput(749.54,352.58)(-0.927,0.492){21}{\rule{0.833pt}{0.119pt}}
\multiput(751.27,351.17)(-20.270,12.000){2}{\rule{0.417pt}{0.400pt}}
\multiput(729.92,348.72)(-0.496,-4.543){41}{\rule{0.120pt}{3.682pt}}
\multiput(730.17,356.36)(-22.000,-189.358){2}{\rule{0.400pt}{1.841pt}}
\multiput(705.77,165.92)(-0.853,-0.493){23}{\rule{0.777pt}{0.119pt}}
\multiput(707.39,166.17)(-20.387,-13.000){2}{\rule{0.388pt}{0.400pt}}
\multiput(682.53,152.93)(-1.252,-0.489){15}{\rule{1.078pt}{0.118pt}}
\multiput(684.76,153.17)(-19.763,-9.000){2}{\rule{0.539pt}{0.400pt}}
\multiput(658.50,143.93)(-1.937,-0.482){9}{\rule{1.567pt}{0.116pt}}
\multiput(661.75,144.17)(-18.748,-6.000){2}{\rule{0.783pt}{0.400pt}}
\multiput(629.85,137.95)(-4.927,-0.447){3}{\rule{3.167pt}{0.108pt}}
\multiput(636.43,138.17)(-16.427,-3.000){2}{\rule{1.583pt}{0.400pt}}
\put(598,134.17){\rule{4.500pt}{0.400pt}}
\multiput(610.66,135.17)(-12.660,-2.000){2}{\rule{2.250pt}{0.400pt}}
\put(576,132.17){\rule{4.500pt}{0.400pt}}
\multiput(588.66,133.17)(-12.660,-2.000){2}{\rule{2.250pt}{0.400pt}}
\put(841.0,133.0){\rule[-0.200pt]{5.300pt}{0.400pt}}
\put(532,130.67){\rule{5.300pt}{0.400pt}}
\multiput(543.00,131.17)(-11.000,-1.000){2}{\rule{2.650pt}{0.400pt}}
\put(554.0,132.0){\rule[-0.200pt]{5.300pt}{0.400pt}}
\put(245.0,131.0){\rule[-0.200pt]{69.138pt}{0.400pt}}
\put(1084,135){\usebox{\plotpoint}}
\put(1062,133.67){\rule{5.300pt}{0.400pt}}
\multiput(1073.00,134.17)(-11.000,-1.000){2}{\rule{2.650pt}{0.400pt}}
\put(929,132.67){\rule{5.541pt}{0.400pt}}
\multiput(940.50,133.17)(-11.500,-1.000){2}{\rule{2.770pt}{0.400pt}}
\put(952.0,134.0){\rule[-0.200pt]{26.499pt}{0.400pt}}
\put(466,133.17){\rule{4.500pt}{0.400pt}}
\multiput(478.66,132.17)(-12.660,2.000){2}{\rule{2.250pt}{0.400pt}}
\multiput(464.92,135.00)(-0.496,0.521){41}{\rule{0.120pt}{0.518pt}}
\multiput(465.17,135.00)(-22.000,21.924){2}{\rule{0.400pt}{0.259pt}}
\multiput(442.92,158.00)(-0.496,3.503){41}{\rule{0.120pt}{2.864pt}}
\multiput(443.17,158.00)(-22.000,146.056){2}{\rule{0.400pt}{1.432pt}}
\multiput(420.92,310.00)(-0.496,1.168){41}{\rule{0.120pt}{1.027pt}}
\multiput(421.17,310.00)(-22.000,48.868){2}{\rule{0.400pt}{0.514pt}}
\multiput(398.92,349.94)(-0.496,-3.249){41}{\rule{0.120pt}{2.664pt}}
\multiput(399.17,355.47)(-22.000,-135.471){2}{\rule{0.400pt}{1.332pt}}
\multiput(376.92,214.08)(-0.496,-1.677){41}{\rule{0.120pt}{1.427pt}}
\multiput(377.17,217.04)(-22.000,-70.038){2}{\rule{0.400pt}{0.714pt}}
\multiput(352.65,145.92)(-0.893,-0.493){23}{\rule{0.808pt}{0.119pt}}
\multiput(354.32,146.17)(-21.324,-13.000){2}{\rule{0.404pt}{0.400pt}}
\multiput(320.41,132.95)(-4.704,-0.447){3}{\rule{3.033pt}{0.108pt}}
\multiput(326.70,133.17)(-15.704,-3.000){2}{\rule{1.517pt}{0.400pt}}
\put(488.0,133.0){\rule[-0.200pt]{106.237pt}{0.400pt}}
\put(245.0,131.0){\rule[-0.200pt]{15.899pt}{0.400pt}}
\put(201.0,131.0){\rule[-0.200pt]{0.400pt}{67.211pt}}
\put(201.0,131.0){\rule[-0.200pt]{212.715pt}{0.400pt}}
\put(1084.0,131.0){\rule[-0.200pt]{0.400pt}{67.211pt}}
\put(201.0,410.0){\rule[-0.200pt]{212.715pt}{0.400pt}}
\end{picture}

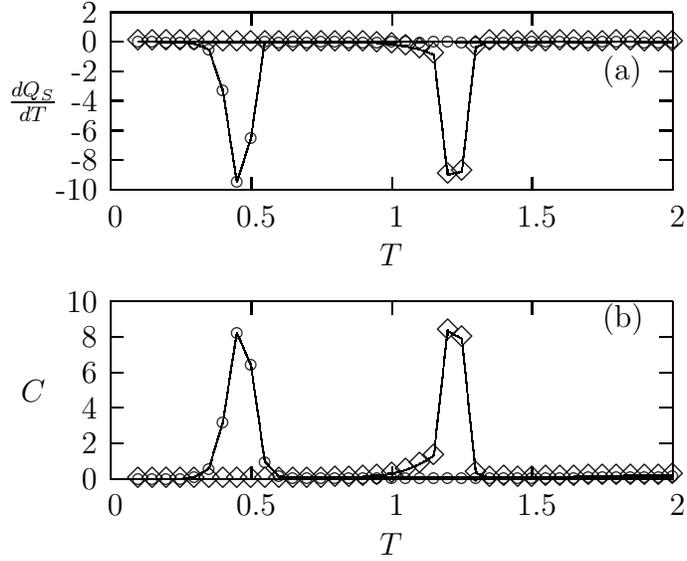
\noindent {\bf Fig-2.} The temperature variations of 
(a) ${{dQ_S} \over {dT}}$
for $h_0= 3.0$ (o) and $h_0=2.0(\diamond)$
and 
(b) $C$ 
for $h_0= 3.0$ (o) and $h_0=2.0$ ($\diamond$).
The continuous lines in all cases joining the data points act
as guides to the eye.
Here, 
$J_A=-0.5J_F$. $T_d(h_0=3.0)=0.45$ and $T_d(h_0=2.0)=1.20$.
\newpage
\setlength{\unitlength}{0.240900pt}
\ifx\plotpoint\undefined\newsavebox{\plotpoint}\fi
\sbox{\plotpoint}{\rule[-0.200pt]{0.400pt}{0.400pt}}%
\begin{picture}(1125,900)(0,0)
\sbox{\plotpoint}{\rule[-0.200pt]{0.400pt}{0.400pt}}%
\put(201.0,131.0){\rule[-0.200pt]{4.818pt}{0.400pt}}
\put(181,131){\makebox(0,0)[r]{-10}}
\put(1064.0,131.0){\rule[-0.200pt]{4.818pt}{0.400pt}}
\put(201.0,253.0){\rule[-0.200pt]{4.818pt}{0.400pt}}
\put(181,253){\makebox(0,0)[r]{-8}}
\put(1064.0,253.0){\rule[-0.200pt]{4.818pt}{0.400pt}}
\put(201.0,374.0){\rule[-0.200pt]{4.818pt}{0.400pt}}
\put(181,374){\makebox(0,0)[r]{-6}}
\put(1064.0,374.0){\rule[-0.200pt]{4.818pt}{0.400pt}}
\put(201.0,496.0){\rule[-0.200pt]{4.818pt}{0.400pt}}
\put(181,496){\makebox(0,0)[r]{-4}}
\put(1064.0,496.0){\rule[-0.200pt]{4.818pt}{0.400pt}}
\put(201.0,617.0){\rule[-0.200pt]{4.818pt}{0.400pt}}
\put(181,617){\makebox(0,0)[r]{-2}}
\put(1064.0,617.0){\rule[-0.200pt]{4.818pt}{0.400pt}}
\put(201.0,739.0){\rule[-0.200pt]{4.818pt}{0.400pt}}
\put(181,739){\makebox(0,0)[r]{ 0}}
\put(1064.0,739.0){\rule[-0.200pt]{4.818pt}{0.400pt}}
\put(201.0,860.0){\rule[-0.200pt]{4.818pt}{0.400pt}}
\put(181,860){\makebox(0,0)[r]{ 2}}
\put(1064.0,860.0){\rule[-0.200pt]{4.818pt}{0.400pt}}
\put(201.0,131.0){\rule[-0.200pt]{0.400pt}{4.818pt}}
\put(201,90){\makebox(0,0){ 0}}
\put(201.0,840.0){\rule[-0.200pt]{0.400pt}{4.818pt}}
\put(422.0,131.0){\rule[-0.200pt]{0.400pt}{4.818pt}}
\put(422,90){\makebox(0,0){ 0.5}}
\put(422.0,840.0){\rule[-0.200pt]{0.400pt}{4.818pt}}
\put(643.0,131.0){\rule[-0.200pt]{0.400pt}{4.818pt}}
\put(643,90){\makebox(0,0){ 1}}
\put(643.0,840.0){\rule[-0.200pt]{0.400pt}{4.818pt}}
\put(863.0,131.0){\rule[-0.200pt]{0.400pt}{4.818pt}}
\put(863,90){\makebox(0,0){ 1.5}}
\put(863.0,840.0){\rule[-0.200pt]{0.400pt}{4.818pt}}
\put(1084.0,131.0){\rule[-0.200pt]{0.400pt}{4.818pt}}
\put(1084,90){\makebox(0,0){ 2}}
\put(1084.0,840.0){\rule[-0.200pt]{0.400pt}{4.818pt}}
\put(201.0,131.0){\rule[-0.200pt]{0.400pt}{175.616pt}}
\put(201.0,131.0){\rule[-0.200pt]{212.715pt}{0.400pt}}
\put(1084.0,131.0){\rule[-0.200pt]{0.400pt}{175.616pt}}
\put(201.0,860.0){\rule[-0.200pt]{212.715pt}{0.400pt}}
\put(80,495){\makebox(0,0){${{dQ_S} \over {dT}}$}}
\put(642,29){\makebox(0,0){$T$}}
\put(1084,739){\raisebox{-.8pt}{\makebox(0,0){$\Diamond$}}}
\put(1062,739){\raisebox{-.8pt}{\makebox(0,0){$\Diamond$}}}
\put(1040,738){\raisebox{-.8pt}{\makebox(0,0){$\Diamond$}}}
\put(1018,739){\raisebox{-.8pt}{\makebox(0,0){$\Diamond$}}}
\put(996,739){\raisebox{-.8pt}{\makebox(0,0){$\Diamond$}}}
\put(974,738){\raisebox{-.8pt}{\makebox(0,0){$\Diamond$}}}
\put(952,731){\raisebox{-.8pt}{\makebox(0,0){$\Diamond$}}}
\put(929,632){\raisebox{-.8pt}{\makebox(0,0){$\Diamond$}}}
\put(907,240){\raisebox{-.8pt}{\makebox(0,0){$\Diamond$}}}
\put(885,280){\raisebox{-.8pt}{\makebox(0,0){$\Diamond$}}}
\put(863,653){\raisebox{-.8pt}{\makebox(0,0){$\Diamond$}}}
\put(841,702){\raisebox{-.8pt}{\makebox(0,0){$\Diamond$}}}
\put(819,725){\raisebox{-.8pt}{\makebox(0,0){$\Diamond$}}}
\put(797,734){\raisebox{-.8pt}{\makebox(0,0){$\Diamond$}}}
\put(775,736){\raisebox{-.8pt}{\makebox(0,0){$\Diamond$}}}
\put(753,738){\raisebox{-.8pt}{\makebox(0,0){$\Diamond$}}}
\put(731,738){\raisebox{-.8pt}{\makebox(0,0){$\Diamond$}}}
\put(709,738){\raisebox{-.8pt}{\makebox(0,0){$\Diamond$}}}
\put(687,738){\raisebox{-.8pt}{\makebox(0,0){$\Diamond$}}}
\put(665,738){\raisebox{-.8pt}{\makebox(0,0){$\Diamond$}}}
\put(643,738){\raisebox{-.8pt}{\makebox(0,0){$\Diamond$}}}
\put(620,738){\raisebox{-.8pt}{\makebox(0,0){$\Diamond$}}}
\put(598,738){\raisebox{-.8pt}{\makebox(0,0){$\Diamond$}}}
\put(576,738){\raisebox{-.8pt}{\makebox(0,0){$\Diamond$}}}
\put(554,738){\raisebox{-.8pt}{\makebox(0,0){$\Diamond$}}}
\put(532,738){\raisebox{-.8pt}{\makebox(0,0){$\Diamond$}}}
\put(510,738){\raisebox{-.8pt}{\makebox(0,0){$\Diamond$}}}
\put(488,738){\raisebox{-.8pt}{\makebox(0,0){$\Diamond$}}}
\put(466,738){\raisebox{-.8pt}{\makebox(0,0){$\Diamond$}}}
\put(444,738){\raisebox{-.8pt}{\makebox(0,0){$\Diamond$}}}
\put(422,738){\raisebox{-.8pt}{\makebox(0,0){$\Diamond$}}}
\put(400,739){\raisebox{-.8pt}{\makebox(0,0){$\Diamond$}}}
\put(378,739){\raisebox{-.8pt}{\makebox(0,0){$\Diamond$}}}
\put(356,739){\raisebox{-.8pt}{\makebox(0,0){$\Diamond$}}}
\put(333,739){\raisebox{-.8pt}{\makebox(0,0){$\Diamond$}}}
\put(311,739){\raisebox{-.8pt}{\makebox(0,0){$\Diamond$}}}
\put(289,739){\raisebox{-.8pt}{\makebox(0,0){$\Diamond$}}}
\put(267,739){\raisebox{-.8pt}{\makebox(0,0){$\Diamond$}}}
\put(245,739){\raisebox{-.8pt}{\makebox(0,0){$\Diamond$}}}
\put(1084,739){\circle{18}}
\put(1062,738){\circle{18}}
\put(1040,738){\circle{18}}
\put(1018,739){\circle{18}}
\put(996,739){\circle{18}}
\put(974,738){\circle{18}}
\put(952,739){\circle{18}}
\put(929,738){\circle{18}}
\put(907,739){\circle{18}}
\put(885,739){\circle{18}}
\put(863,738){\circle{18}}
\put(841,738){\circle{18}}
\put(819,739){\circle{18}}
\put(797,738){\circle{18}}
\put(775,738){\circle{18}}
\put(753,738){\circle{18}}
\put(731,738){\circle{18}}
\put(709,739){\circle{18}}
\put(687,738){\circle{18}}
\put(665,736){\circle{18}}
\put(643,735){\circle{18}}
\put(620,692){\circle{18}}
\put(598,671){\circle{18}}
\put(576,211){\circle{18}}
\put(554,210){\circle{18}}
\put(532,707){\circle{18}}
\put(510,731){\circle{18}}
\put(488,738){\circle{18}}
\put(466,738){\circle{18}}
\put(444,738){\circle{18}}
\put(422,738){\circle{18}}
\put(400,738){\circle{18}}
\put(378,738){\circle{18}}
\put(356,738){\circle{18}}
\put(333,739){\circle{18}}
\put(311,738){\circle{18}}
\put(289,739){\circle{18}}
\put(267,739){\circle{18}}
\put(245,739){\circle{18}}
\put(1084,739){\usebox{\plotpoint}}
\put(1040,737.67){\rule{5.300pt}{0.400pt}}
\multiput(1051.00,738.17)(-11.000,-1.000){2}{\rule{2.650pt}{0.400pt}}
\put(1018,737.67){\rule{5.300pt}{0.400pt}}
\multiput(1029.00,737.17)(-11.000,1.000){2}{\rule{2.650pt}{0.400pt}}
\put(1062.0,739.0){\rule[-0.200pt]{5.300pt}{0.400pt}}
\put(974,737.67){\rule{5.300pt}{0.400pt}}
\multiput(985.00,738.17)(-11.000,-1.000){2}{\rule{2.650pt}{0.400pt}}
\multiput(968.37,736.93)(-1.637,-0.485){11}{\rule{1.357pt}{0.117pt}}
\multiput(971.18,737.17)(-19.183,-7.000){2}{\rule{0.679pt}{0.400pt}}
\multiput(950.92,723.44)(-0.496,-2.177){43}{\rule{0.120pt}{1.822pt}}
\multiput(951.17,727.22)(-23.000,-95.219){2}{\rule{0.400pt}{0.911pt}}
\multiput(927.92,602.00)(-0.496,-9.050){41}{\rule{0.120pt}{7.227pt}}
\multiput(928.17,617.00)(-22.000,-376.999){2}{\rule{0.400pt}{3.614pt}}
\multiput(905.92,240.00)(-0.496,0.914){41}{\rule{0.120pt}{0.827pt}}
\multiput(906.17,240.00)(-22.000,38.283){2}{\rule{0.400pt}{0.414pt}}
\multiput(883.92,280.00)(-0.496,8.611){41}{\rule{0.120pt}{6.882pt}}
\multiput(884.17,280.00)(-22.000,358.716){2}{\rule{0.400pt}{3.441pt}}
\multiput(861.92,653.00)(-0.496,1.122){41}{\rule{0.120pt}{0.991pt}}
\multiput(862.17,653.00)(-22.000,46.943){2}{\rule{0.400pt}{0.495pt}}
\multiput(839.92,702.00)(-0.496,0.521){41}{\rule{0.120pt}{0.518pt}}
\multiput(840.17,702.00)(-22.000,21.924){2}{\rule{0.400pt}{0.259pt}}
\multiput(814.53,725.59)(-1.252,0.489){15}{\rule{1.078pt}{0.118pt}}
\multiput(816.76,724.17)(-19.763,9.000){2}{\rule{0.539pt}{0.400pt}}
\put(775,734.17){\rule{4.500pt}{0.400pt}}
\multiput(787.66,733.17)(-12.660,2.000){2}{\rule{2.250pt}{0.400pt}}
\put(753,736.17){\rule{4.500pt}{0.400pt}}
\multiput(765.66,735.17)(-12.660,2.000){2}{\rule{2.250pt}{0.400pt}}
\put(996.0,739.0){\rule[-0.200pt]{5.300pt}{0.400pt}}
\put(400,737.67){\rule{5.300pt}{0.400pt}}
\multiput(411.00,737.17)(-11.000,1.000){2}{\rule{2.650pt}{0.400pt}}
\put(422.0,738.0){\rule[-0.200pt]{79.738pt}{0.400pt}}
\put(245.0,739.0){\rule[-0.200pt]{37.339pt}{0.400pt}}
\put(1084,739){\usebox{\plotpoint}}
\put(1062,737.67){\rule{5.300pt}{0.400pt}}
\multiput(1073.00,738.17)(-11.000,-1.000){2}{\rule{2.650pt}{0.400pt}}
\put(1018,737.67){\rule{5.300pt}{0.400pt}}
\multiput(1029.00,737.17)(-11.000,1.000){2}{\rule{2.650pt}{0.400pt}}
\put(1040.0,738.0){\rule[-0.200pt]{5.300pt}{0.400pt}}
\put(974,737.67){\rule{5.300pt}{0.400pt}}
\multiput(985.00,738.17)(-11.000,-1.000){2}{\rule{2.650pt}{0.400pt}}
\put(952,737.67){\rule{5.300pt}{0.400pt}}
\multiput(963.00,737.17)(-11.000,1.000){2}{\rule{2.650pt}{0.400pt}}
\put(929,737.67){\rule{5.541pt}{0.400pt}}
\multiput(940.50,738.17)(-11.500,-1.000){2}{\rule{2.770pt}{0.400pt}}
\put(907,737.67){\rule{5.300pt}{0.400pt}}
\multiput(918.00,737.17)(-11.000,1.000){2}{\rule{2.650pt}{0.400pt}}
\put(996.0,739.0){\rule[-0.200pt]{5.300pt}{0.400pt}}
\put(863,737.67){\rule{5.300pt}{0.400pt}}
\multiput(874.00,738.17)(-11.000,-1.000){2}{\rule{2.650pt}{0.400pt}}
\put(885.0,739.0){\rule[-0.200pt]{5.300pt}{0.400pt}}
\put(819,737.67){\rule{5.300pt}{0.400pt}}
\multiput(830.00,737.17)(-11.000,1.000){2}{\rule{2.650pt}{0.400pt}}
\put(797,737.67){\rule{5.300pt}{0.400pt}}
\multiput(808.00,738.17)(-11.000,-1.000){2}{\rule{2.650pt}{0.400pt}}
\put(841.0,738.0){\rule[-0.200pt]{5.300pt}{0.400pt}}
\put(709,737.67){\rule{5.300pt}{0.400pt}}
\multiput(720.00,737.17)(-11.000,1.000){2}{\rule{2.650pt}{0.400pt}}
\put(687,737.67){\rule{5.300pt}{0.400pt}}
\multiput(698.00,738.17)(-11.000,-1.000){2}{\rule{2.650pt}{0.400pt}}
\put(665,736.17){\rule{4.500pt}{0.400pt}}
\multiput(677.66,737.17)(-12.660,-2.000){2}{\rule{2.250pt}{0.400pt}}
\put(643,734.67){\rule{5.300pt}{0.400pt}}
\multiput(654.00,735.17)(-11.000,-1.000){2}{\rule{2.650pt}{0.400pt}}
\multiput(641.92,731.48)(-0.496,-0.940){43}{\rule{0.120pt}{0.848pt}}
\multiput(642.17,733.24)(-23.000,-41.240){2}{\rule{0.400pt}{0.424pt}}
\multiput(617.85,690.92)(-0.522,-0.496){39}{\rule{0.519pt}{0.119pt}}
\multiput(618.92,691.17)(-20.923,-21.000){2}{\rule{0.260pt}{0.400pt}}
\multiput(596.92,635.87)(-0.496,-10.622){41}{\rule{0.120pt}{8.464pt}}
\multiput(597.17,653.43)(-22.000,-442.433){2}{\rule{0.400pt}{4.232pt}}
\put(554,209.67){\rule{5.300pt}{0.400pt}}
\multiput(565.00,210.17)(-11.000,-1.000){2}{\rule{2.650pt}{0.400pt}}
\multiput(552.92,210.00)(-0.496,11.477){41}{\rule{0.120pt}{9.136pt}}
\multiput(553.17,210.00)(-22.000,478.037){2}{\rule{0.400pt}{4.568pt}}
\multiput(530.92,707.00)(-0.496,0.544){41}{\rule{0.120pt}{0.536pt}}
\multiput(531.17,707.00)(-22.000,22.887){2}{\rule{0.400pt}{0.268pt}}
\multiput(504.37,731.59)(-1.637,0.485){11}{\rule{1.357pt}{0.117pt}}
\multiput(507.18,730.17)(-19.183,7.000){2}{\rule{0.679pt}{0.400pt}}
\put(731.0,738.0){\rule[-0.200pt]{15.899pt}{0.400pt}}
\put(333,737.67){\rule{5.541pt}{0.400pt}}
\multiput(344.50,737.17)(-11.500,1.000){2}{\rule{2.770pt}{0.400pt}}
\put(311,737.67){\rule{5.300pt}{0.400pt}}
\multiput(322.00,738.17)(-11.000,-1.000){2}{\rule{2.650pt}{0.400pt}}
\put(289,737.67){\rule{5.300pt}{0.400pt}}
\multiput(300.00,737.17)(-11.000,1.000){2}{\rule{2.650pt}{0.400pt}}
\put(356.0,738.0){\rule[-0.200pt]{31.799pt}{0.400pt}}
\put(245.0,739.0){\rule[-0.200pt]{10.600pt}{0.400pt}}
\put(201.0,131.0){\rule[-0.200pt]{0.400pt}{175.616pt}}
\put(201.0,131.0){\rule[-0.200pt]{212.715pt}{0.400pt}}
\put(1084.0,131.0){\rule[-0.200pt]{0.400pt}{175.616pt}}
\put(201.0,860.0){\rule[-0.200pt]{212.715pt}{0.400pt}}
\end{picture}

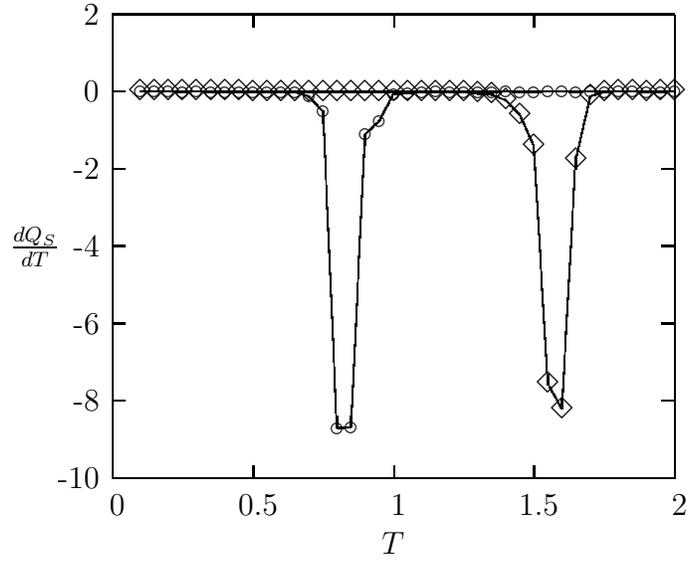
\noindent {\bf Fig-3.} The temperature variations of 
${{dQ_S} \over {dT}}$ for
$h_0=2.0(T_d=1.6)(\diamond)$ and $h_0=3.0(T_d=0.8)$(o) 
for pure ferromagnetic ($J_F=J_A=1.0$) case.
\newpage
\setlength{\unitlength}{0.240900pt}
\ifx\plotpoint\undefined\newsavebox{\plotpoint}\fi
\sbox{\plotpoint}{\rule[-0.200pt]{0.400pt}{0.400pt}}%
\begin{picture}(1125,900)(0,0)
\sbox{\plotpoint}{\rule[-0.200pt]{0.400pt}{0.400pt}}%
\put(221.0,131.0){\rule[-0.200pt]{4.818pt}{0.400pt}}
\put(201,131){\makebox(0,0)[r]{ 0}}
\put(1064.0,131.0){\rule[-0.200pt]{4.818pt}{0.400pt}}
\put(221.0,212.0){\rule[-0.200pt]{4.818pt}{0.400pt}}
\put(201,212){\makebox(0,0)[r]{ 0.5}}
\put(1064.0,212.0){\rule[-0.200pt]{4.818pt}{0.400pt}}
\put(221.0,293.0){\rule[-0.200pt]{4.818pt}{0.400pt}}
\put(201,293){\makebox(0,0)[r]{ 1}}
\put(1064.0,293.0){\rule[-0.200pt]{4.818pt}{0.400pt}}
\put(221.0,374.0){\rule[-0.200pt]{4.818pt}{0.400pt}}
\put(201,374){\makebox(0,0)[r]{ 1.5}}
\put(1064.0,374.0){\rule[-0.200pt]{4.818pt}{0.400pt}}
\put(221.0,455.0){\rule[-0.200pt]{4.818pt}{0.400pt}}
\put(201,455){\makebox(0,0)[r]{ 2}}
\put(1064.0,455.0){\rule[-0.200pt]{4.818pt}{0.400pt}}
\put(221.0,536.0){\rule[-0.200pt]{4.818pt}{0.400pt}}
\put(201,536){\makebox(0,0)[r]{ 2.5}}
\put(1064.0,536.0){\rule[-0.200pt]{4.818pt}{0.400pt}}
\put(221.0,617.0){\rule[-0.200pt]{4.818pt}{0.400pt}}
\put(201,617){\makebox(0,0)[r]{ 3}}
\put(1064.0,617.0){\rule[-0.200pt]{4.818pt}{0.400pt}}
\put(221.0,698.0){\rule[-0.200pt]{4.818pt}{0.400pt}}
\put(201,698){\makebox(0,0)[r]{ 3.5}}
\put(1064.0,698.0){\rule[-0.200pt]{4.818pt}{0.400pt}}
\put(221.0,779.0){\rule[-0.200pt]{4.818pt}{0.400pt}}
\put(201,779){\makebox(0,0)[r]{ 4}}
\put(1064.0,779.0){\rule[-0.200pt]{4.818pt}{0.400pt}}
\put(221.0,860.0){\rule[-0.200pt]{4.818pt}{0.400pt}}
\put(201,860){\makebox(0,0)[r]{ 4.5}}
\put(1064.0,860.0){\rule[-0.200pt]{4.818pt}{0.400pt}}
\put(221.0,131.0){\rule[-0.200pt]{0.400pt}{4.818pt}}
\put(221,90){\makebox(0,0){ 0}}
\put(221.0,840.0){\rule[-0.200pt]{0.400pt}{4.818pt}}
\put(317.0,131.0){\rule[-0.200pt]{0.400pt}{4.818pt}}
\put(317,90){\makebox(0,0){ 0.5}}
\put(317.0,840.0){\rule[-0.200pt]{0.400pt}{4.818pt}}
\put(413.0,131.0){\rule[-0.200pt]{0.400pt}{4.818pt}}
\put(413,90){\makebox(0,0){ 1}}
\put(413.0,840.0){\rule[-0.200pt]{0.400pt}{4.818pt}}
\put(509.0,131.0){\rule[-0.200pt]{0.400pt}{4.818pt}}
\put(509,90){\makebox(0,0){ 1.5}}
\put(509.0,840.0){\rule[-0.200pt]{0.400pt}{4.818pt}}
\put(605.0,131.0){\rule[-0.200pt]{0.400pt}{4.818pt}}
\put(605,90){\makebox(0,0){ 2}}
\put(605.0,840.0){\rule[-0.200pt]{0.400pt}{4.818pt}}
\put(700.0,131.0){\rule[-0.200pt]{0.400pt}{4.818pt}}
\put(700,90){\makebox(0,0){ 2.5}}
\put(700.0,840.0){\rule[-0.200pt]{0.400pt}{4.818pt}}
\put(796.0,131.0){\rule[-0.200pt]{0.400pt}{4.818pt}}
\put(796,90){\makebox(0,0){ 3}}
\put(796.0,840.0){\rule[-0.200pt]{0.400pt}{4.818pt}}
\put(892.0,131.0){\rule[-0.200pt]{0.400pt}{4.818pt}}
\put(892,90){\makebox(0,0){ 3.5}}
\put(892.0,840.0){\rule[-0.200pt]{0.400pt}{4.818pt}}
\put(988.0,131.0){\rule[-0.200pt]{0.400pt}{4.818pt}}
\put(988,90){\makebox(0,0){ 4}}
\put(988.0,840.0){\rule[-0.200pt]{0.400pt}{4.818pt}}
\put(1084.0,131.0){\rule[-0.200pt]{0.400pt}{4.818pt}}
\put(1084,90){\makebox(0,0){ 4.5}}
\put(1084.0,840.0){\rule[-0.200pt]{0.400pt}{4.818pt}}
\put(221.0,131.0){\rule[-0.200pt]{0.400pt}{175.616pt}}
\put(221.0,131.0){\rule[-0.200pt]{207.897pt}{0.400pt}}
\put(1084.0,131.0){\rule[-0.200pt]{0.400pt}{175.616pt}}
\put(221.0,860.0){\rule[-0.200pt]{207.897pt}{0.400pt}}
\put(80,495){\makebox(0,0){$h_0$}}
\put(652,29){\makebox(0,0){$T$}}
\put(288,682){\raisebox{-.8pt}{\makebox(0,0){$\Diamond$}}}
\put(307,617){\raisebox{-.8pt}{\makebox(0,0){$\Diamond$}}}
\put(336,585){\raisebox{-.8pt}{\makebox(0,0){$\Diamond$}}}
\put(374,536){\raisebox{-.8pt}{\makebox(0,0){$\Diamond$}}}
\put(403,504){\raisebox{-.8pt}{\makebox(0,0){$\Diamond$}}}
\put(422,487){\raisebox{-.8pt}{\makebox(0,0){$\Diamond$}}}
\put(432,471){\raisebox{-.8pt}{\makebox(0,0){$\Diamond$}}}
\put(451,455){\raisebox{-.8pt}{\makebox(0,0){$\Diamond$}}}
\put(480,439){\raisebox{-.8pt}{\makebox(0,0){$\Diamond$}}}
\put(499,423){\raisebox{-.8pt}{\makebox(0,0){$\Diamond$}}}
\put(280,200){\it Ordered phase}
\put(528,406){\raisebox{-.8pt}{\makebox(0,0){$\Diamond$}}}
\put(557,390){\raisebox{-.8pt}{\makebox(0,0){$\Diamond$}}}
\put(585,374){\raisebox{-.8pt}{\makebox(0,0){$\Diamond$}}}
\put(624,358){\raisebox{-.8pt}{\makebox(0,0){$\Diamond$}}}
\put(672,342){\raisebox{-.8pt}{\makebox(0,0){$\Diamond$}}}
\put(720,325){\raisebox{-.8pt}{\makebox(0,0){$\Diamond$}}}
\put(768,309){\raisebox{-.8pt}{\makebox(0,0){$\Diamond$}}}
\put(816,293){\raisebox{-.8pt}{\makebox(0,0){$\Diamond$}}}
\put(844,277){\raisebox{-.8pt}{\makebox(0,0){$\Diamond$}}}
\put(863,261){\raisebox{-.8pt}{\makebox(0,0){$\Diamond$}}}
\put(873,244){\raisebox{-.8pt}{\makebox(0,0){$\Diamond$}}}
\put(892,228){\raisebox{-.8pt}{\makebox(0,0){$\Diamond$}}}
\put(911,147){\raisebox{-.8pt}{\makebox(0,0){$\Diamond$}}}
\put(317,779){\circle{18}}
\put(326,763){\circle{18}}
\put(346,747){\circle{18}}
\put(355,730){\circle{18}}
\put(374,714){\circle{18}}
\put(384,698){\circle{18}}
\put(403,682){\circle{18}}
\put(422,666){\circle{18}}
\put(442,649){\circle{18}}
\put(461,633){\circle{18}}
\put(480,617){\circle{18}}
\put(509,601){\circle{18}}
\put(537,585){\circle{18}}
\put(566,568){\circle{18}}
\put(595,552){\circle{18}}
\put(643,536){\circle{18}}
\put(681,520){\circle{18}}
\put(739,504){\circle{18}}
\put(645,710){\it Disordered phase}
\put(796,487){\circle{18}}
\put(844,471){\circle{18}}
\put(892,455){\circle{18}}
\put(921,439){\circle{18}}
\put(950,423){\circle{18}}
\put(969,406){\circle{18}}
\put(988,390){\circle{18}}
\put(998,374){\circle{18}}
\put(1017,358){\circle{18}}
\put(1026,342){\circle{18}}
\put(1026,325){\circle{18}}
\put(1046,309){\circle{18}}
\put(1046,293){\circle{18}}
\put(1055,277){\circle{18}}
\put(1055,261){\circle{18}}
\put(1065,244){\circle{18}}
\put(1065,228){\circle{18}}
\put(1065,212){\circle{18}}
\put(1065,196){\circle{18}}
\put(1074,180){\circle{18}}
\put(1074,163){\circle{18}}
\put(1074,147){\circle{18}}
\sbox{\plotpoint}{\rule[-0.500pt]{1.000pt}{1.000pt}}%
\put(288,779){\circle*{24}}
\put(298,763){\circle*{24}}
\put(307,747){\circle*{24}}
\put(317,730){\circle*{24}}
\put(317,714){\circle*{24}}
\put(326,698){\circle*{24}}
\put(336,682){\circle*{24}}
\put(355,666){\circle*{24}}
\put(365,649){\circle*{24}}
\put(374,633){\circle*{24}}
\put(384,617){\circle*{24}}
\put(394,601){\circle*{24}}
\put(403,585){\circle*{24}}
\put(413,568){\circle*{24}}
\put(432,552){\circle*{24}}
\put(442,536){\circle*{24}}
\put(461,520){\circle*{24}}
\put(470,504){\circle*{24}}
\put(489,487){\circle*{24}}
\put(499,471){\circle*{24}}
\put(518,455){\circle*{24}}
\put(537,439){\circle*{24}}
\put(557,423){\circle*{24}}
\put(576,406){\circle*{24}}
\put(605,390){\circle*{24}}
\put(624,374){\circle*{24}}
\put(643,358){\circle*{24}}
\put(672,342){\circle*{24}}
\put(691,325){\circle*{24}}
\put(720,309){\circle*{24}}
\put(739,293){\circle*{24}}
\put(768,277){\circle*{24}}
\put(796,261){\circle*{24}}
\put(825,244){\circle*{24}}
\put(854,228){\circle*{24}}
\put(883,212){\circle*{24}}
\put(921,196){\circle*{24}}
\put(969,180){\circle*{24}}
\put(1007,163){\circle*{24}}
\sbox{\plotpoint}{\rule[-0.200pt]{0.400pt}{0.400pt}}%
\put(221.0,131.0){\rule[-0.200pt]{0.400pt}{175.616pt}}
\put(221.0,131.0){\rule[-0.200pt]{207.897pt}{0.400pt}}
\put(1084.0,131.0){\rule[-0.200pt]{0.400pt}{175.616pt}}
\put(221.0,860.0){\rule[-0.200pt]{207.897pt}{0.400pt}}
\end{picture}

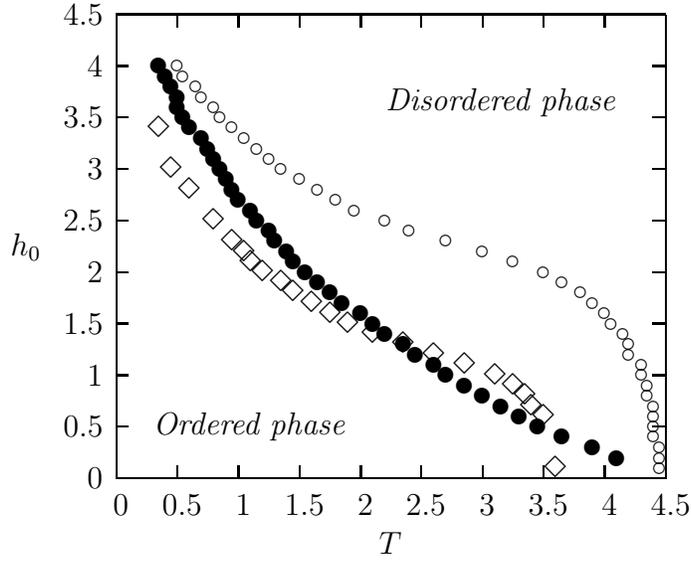
\noindent {\bf Fig-4.} The phase diagram of dynamic phase transitions.
Metamagnetic dynamic phase bounadries (i) $(\diamond)$ for $J_A=-0.5$ 
and $J_F=1.0$ and (ii) (o) for $J_A=-1.0$ and $J_F=1.0$. The ($\bullet$)
represents the dynamic phase bounadry of pure ising ferromagnets
($J_A=J_F=1.0$).
\newpage
\end{document}